\documentclass{aastex62}
\usepackage{longtable}
\usepackage{multirow}
\usepackage{graphicx}           
\usepackage{amsmath,amssymb}
\usepackage[T1]{fontenc}

\graphicspath{{./}{figures/}}

\submitjournal{ApJ}

\shorttitle{iMOGABA: Mrk\,421}
\shortauthors{Lee et al.}

\begin{document}
\title{A study of the radio spectrum of Mrk\,421}
\correspondingauthor{Sang-Sung Lee}
\email{sslee@kasi.re.kr}

\author[0000-0002-0786-7307]{Jee Won Lee}
\affiliation{Korea Astronomy and Space Science Institute, 776 Daedeok-daero, Yuseong-gu, Daejeon 34055, Republic of Korea}

\author[0000-0002-6269-594X]{Sang-Sung Lee}
\affiliation{Korea Astronomy and Space Science Institute, 776 Daedeok-daero, Yuseong-gu, Daejeon 34055, Republic of Korea}
\affiliation{University of Science and Technology, 217 Gajeong-ro, Yuseong-gu, Daejeon 34113, Korea}

\author[0000-0001-6094-9291]{Jeffrey Hodgson}
\affil{Department of Physics and Astronomy, Sejong University, 209 Neungdong-ro, Gwangjin-gu, Seoul, South Korea}

\author[0000-0001-6993-1696]{Algaba Juan-Carlos}
\affiliation{Department of Physics, Faculty of Science, Universiti Malaya, 50603 Kuala Lumpur, Malaysia}

\author[0000-0001-7556-8504]{Sang-Hyun Kim}
\affiliation{Korea Astronomy and Space Science Institute, 776 Daedeok-daero, Yuseong-gu, Daejeon 34055, Republic of Korea}
\affiliation{University of Science and Technology, 217 Gajeong-ro, Yuseong-gu, Daejeon 34113, Korea}

\author[0009-0002-1871-5824]{Whee Yeon Cheong}
\affiliation{Korea Astronomy and Space Science Institute, 776 Daedeok-daero, Yuseong-gu, Daejeon 34055, Republic of Korea}
\affiliation{University of Science and Technology, 217 Gajeong-ro, Yuseong-gu, Daejeon 34113, Korea}

\author[0009-0005-7629-8450]{Hyeon-Woo Jeong}
\affiliation{Korea Astronomy and Space Science Institute, 776 Daedeok-daero, Yuseong-gu, Daejeon 34055, Republic of Korea}
\affiliation{University of Science and Technology, 217 Gajeong-ro, Yuseong-gu, Daejeon 34113, Korea}

\author[0000-0002-0112-4836]{Sincheol Kang}
\affiliation{Korea Astronomy and Space Science Institute, 776 Daedeok-daero, Yuseong-gu, Daejeon 34055, Republic of Korea}

\begin{abstract}
We present the results of a spectral analysis using simultaneous multifrequency (22, 43, 86, and 129~GHz) very long baseline interferometry (VLBI) observations of the Korean VLBI Network (KVN) on BL Lac object, Markarian\,421 (Mrk\,421). The data we used was obtained from January 2013 to June 2018. 
The light curves showed several flux enhancements with global decreases.
To separate the variable and quiescent components in the multifrequency light curves for milliarcsecond-scale emission regions, we assumed that the quiescent radiation comes from the emission regions radiating constant optically-thin synchrotron emissions (i.e., a minimum flux density with an optically thin spectral index). The quiescent spectrum determined from the multifrequency light curves was subtracted from the total CLEAN flux density, yielding a variable component in the flux that produces the time-dependent spectrum. 
We found that the observed spectra were flat at 22-43\,GHz, and relatively steep at 43-86~GHz, whereas the quiescent-corrected spectra are sometimes quite different from the observed spectra (e.g., sometimes inverted at 22-43\,GHz ).
The quiescent-corrected spectral indices were much more variable than the observed spectral indices.
This spectral investigation implies that the quiescent-spectrum correction can significantly affect the multifrequency spectral index of variable compact radio sources such as blazars. Therefore, the synchrotron self-absorption B-field strength $(B_{\rm SSA})$ can be significantly affected because $B_{\rm SSA}$ is proportional to the fifth power of turnover frequency. 
\end{abstract}


\section{Introduction}\label{sec:intro}
Active Galactic Nuclei (AGN) are the central regions of galaxies that exhibit unusually bright luminosity across multiwavelengths, from radio to gamma rays. 
It is thought that the mechanism responsible for this luminosity is a supermassive black hole ($10^{6} - 10^{10}\,\rm M_{\odot})$ at the center of the AGN, which accretes enormous amounts of matter from its surrounding environment. A relativistic jet is formed by the accreted matter \citep{Ulrich+97}.  

A blazar is a subclass of AGN that exhibits extremely energetic and highly variable observational behavior. 
The source is among the most luminous and powerful objects, and features a relativistic jet that points directly toward us with a very small viewing angle. This alignment causes a blazar to appear as a point-like source in the sky that exhibits rapid and dramatic variability in its brightness \citep{Urry+95}.
Blazars encompass both BL Lac objects and flat-spectrum radio quasars (FSRQs).
While FSRQs and BL Lac objects share similar characteristics, such as being highly variable and bright, their spectra differ. The spectrum of a BL Lac object has weak or no broad emission, unlike that of an FSRQ.

Markarian\,421 (Mrk\,421), is one of the most extensively studied high-synchrotron-peaked blazars and one of the closest sources, located at a redshift z=0.031. Its high luminosity enables us to investigate its emission \citep{Katarzynski+03,Charlot+06,Horan+09,Acciari+11,Aleksic+15,Arbet-Engels+21}. Its source Mrk\,421 was the first extragalactic source detected in the TeV energy bands, as reported by \citet{Punch+92} using the Whipple 10-m Cherenkov telescope. Mrk\,421 is classified as a BL Lac object.

\citet{Acciari+11} reported no correlation between TeV $\gamma$-ray and optical/radio light curves obtained from observations performed from December 2005 to May 2008 with VERITAS/Whipple gamma-ray data and Metsahovi and UMRAO radio data. 
On the other hand, when focusing on the very high energy (VHE) E > 100 GeV, they found a strong correlation between TeV gamma-ray and X-rays and suggested that this can be explained by the one-zone synchrotron self-Compton model \citep{Zhu+16}.
In a multiwavelength study conducted on Mrk\,421 over 5.5-yr period, from December 2012 to April 2018, \citet{Arbet-Engels+21} performed several cross-correlation analyses. They found that the TeV and X-ray light curves were very well correlated, with a lag of less than 0.6 days. The GeV and 15\,GHz radio light curves also showed correlation, with a lag of over 30-100 days. However, no correlation was observed between the TeV and the radio light curves. They suggested that the variability at high energy may be due to the leptonic mechanism, given the short time scale of variability.

Although extensive multiwavelength studies on Mrk\,421 have been extensively conducted to date, research on its radio spectrum has been limited. In particular, simultaneous multifrequency very long baseline interferometry (VLBI) imaging observations with high time resolution have not been performed yet. In this study, we present various analyses of the radio spectrum of Mrk\,421 to explain the correlation between the spectrum and the flux density variations.

In this paper, we report results from simultaneous multifrequency VLBI observations of Mrk\,421 carried out using the Korea VLBI Network (KVN) at 22, 43, and 86, from January 2013 to June 2018. The paper is organized as follows. In Section \ref{sec:obs}, we present descriptions of the observations for both the KVN and all archival data used in this paper. Section \ref{sec:result} and \ref{sec:analysis} describe the results and various analyses of flux density variability at multiple frequencies, respectively. Finally, Section \ref{sec:discussion} includes a discussion on the results and analysis.

\begin{figure*}
\centering
\includegraphics[angle=270, width=1.\textwidth]{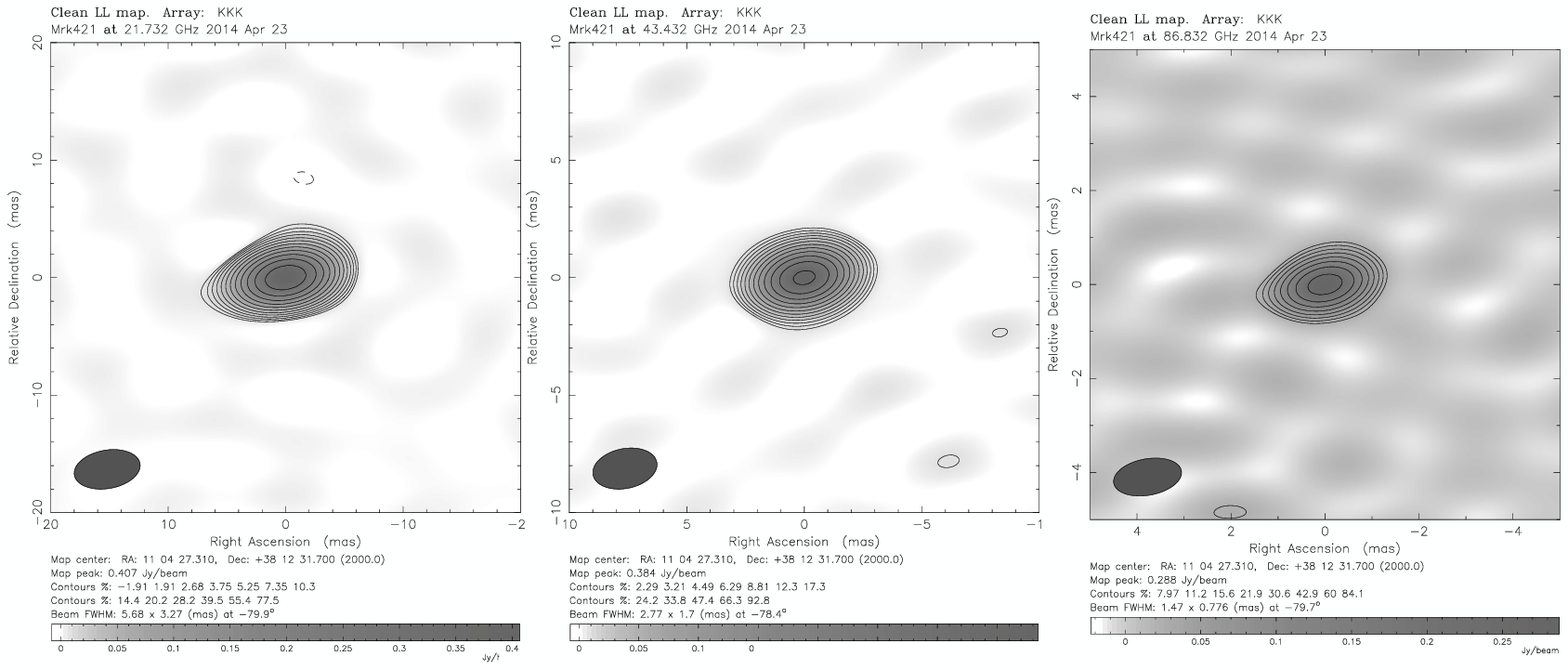}
\caption{\label{fig:clean_map}CLEAN maps of Mrk\,421 obtained in Epoch 14 (April 23, 2014). The left, middle, and right panels indicate source structures at 22, 43, and 86\,GHz, respectively. The source shows a compact core-dominated structure at all frequencies. The contour level starts from 3 times the noise and increases by a factor of $\sqrt{2}$. The beam size is indicated by the filled ellipse in the bottom left corner.}
\end{figure*}
\section{Observations and Data calibration}\label{sec:obs}
\subsection{KVN observations}
We conducted simultaneous multifrequency VLBI imaging monitoring observations of Mrk\,421 using the Korean VLBI Network (KVN) at 22, 43, 86, and 129\,GHz as part of the iMOGABA(Interferometric Monitoring of Gamma-ray Bright active galactic Nuclei) project \citep{Lee+16,Lee+17b,Algaba+18a,Lee+20,Kang+21}.
The KVN is a VLBI network system consisting of three 21-meter radio telescopes located in Seoul (KVN Yonsei), Ulsan (KVN Ulsan), and Jeju (KVN Tamna), Korea.
The monitoring observations were primarily carried out every month.
However, we excluded the 129\,GHz data from the analysis because there were no valid measurements, higher than 3-$\sigma$ of errors.  
The observation period was from January 2013 to June 2018, except for maintenance periods from June to August.
Measurements were made in 41 epochs over a period of 5.5 years.
At 22, 43, and 86\,GHz, 40 epochs, 33 epochs, and 21 epochs were measured, respectively. 
The observing frequencies were 21.700-21.764, 43.400-43.464, 86.800-86.864, and 129.300-129.364\,GHz, with a total bandwidth of 256\,MHz. 
The angular resolutions of the KVN were 6, 3, and 1.5 mas at 22, 43, and 86\,GHz, respectively.
We used a left circular polarization observation mode at a recording rate of 1 Gbps. 
For more details of the KVN observations, see \cite{Lee+16,Lee+17a}. 

\begin{figure*}
\centering
\includegraphics[width=1.\textwidth]{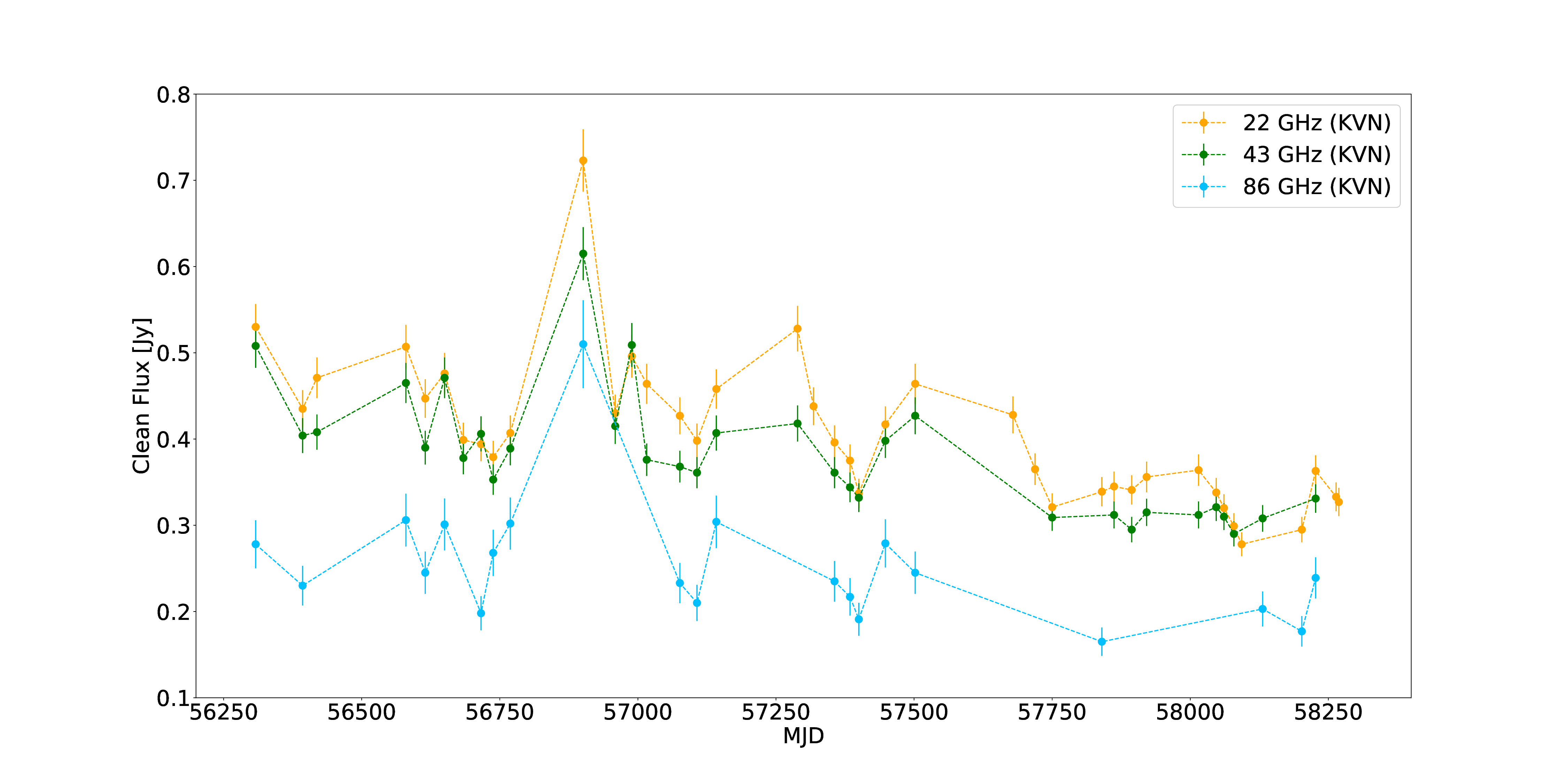}
\caption{\label{fig:lc} Light curves of Mrk\,421. Yellow, green, and sky blue dots indicate 22, 43, and 86\,GHz, respectively. }
\end{figure*}

\subsection{Data calibration}

The observed data were processed using the DiFX software correlator in Daejeon, Korea \citep{Deller+07,Lee+15}. 
The DiFX correlator generated a cross-correlation function spectrum with a resolution of 0.125 MHz and an accumulation period of one second.
After the correlation, the data was calibrated with the AIPS (Astronomy Image Processing System) software package, provided by the National Radio Astronomy Observatory (NRAO), following standard procedures, including phase and amplitude calibration, fringe fitting, and bandpass calibration.
We used the KVN pipeline developed by \citet{Hodgson+16} to perform the data calibration.
Sensitivity at high frequencies was improved by transferring phase solutions from lower to higher frequencies (i.e., FPT) as described by \citet{Rioja+11} and \citet{Algaba+15}.
Amplitude calibration was conducted using the system temperatures measured during the observations. 
The system temperatures were adjusted for atmospheric opacity based on sky-tipping curve measurements conducted every hour at each radio telescope.
Renormalization of the fringe amplitudes was conducted to correct the amplitude distortion due to quantization, and losses from quantization and re-quantization were corrected \citep{Lee+15}. 
After the amplitude calibration and the re-quantization loss calibration, the uncertainty of the amplitude calibration was within 5\,\% at 22 and 43\,GHz and 10-30\,\% at 86 and 129\,GHz. 

\subsection{Imaging and modelfitting}
After the phase and amplitude calibration, we used DIFMAP to make a CLEAN contour map.
To make a CLEAN map with the phase self-calibration, for the first time we performed the startmod command. This uses a point source model of 1\,Jy to conduct the initial phase self-calibration.
Then, the visibility data at 22, 43, and 86 GHz were averaged at 30-second intervals. This averaging time aligned with the typical coherence timescales of KVN observations.
We flagged bad amplitude and phase data points and outliers in the vplot.
Then, the repetitive CLEANing and phase self-calibration within the central emitting regions were performed until the noise root-mean-square (rms) level was no longer significantly lowered. The standard CLEANing and self-calibration within the central emitting regions were conducted until no significant flux density was added compared to the image rms level.

Since Mrk\,421 is a compact source observed on mas-scales in the KVN observations, finding the best model is straightforward.
Figure\,\ref{fig:clean_map} displays the CLEAN contour maps of Mrk\,421 at 22, 43, and 86\,GHz for a representative epoch (well observed at all frequencies) showing compact core-dominated components.
After making the CLEAN contour map, we computed the image quality factor $\xi_{\rm r}$. This is the ratio of the image rms noise and its mathematical expectation, $\xi_{\rm r} = S_{\rm r}/S_{\rm r,exp}$, where $S_{\rm r}$ is the maximum absolute flux density in the residual image, and $S_{\rm r,exp}$ is the expectation of $S_{\rm r}$. 
For a more comprehensive understanding of this image quality evaluation scheme, please refer to \citet{Lobanov+06} and \citet{Lee+16,Lee+17a}.
The $\xi_{\rm r}$ values derived in this paper range from 0.65 - 0.97, suggesting that the images sufficiently represent the structure identified in the visibility data.

To establish the model of the source, we fit the model in the \textit{uv}-data.
In this process, we used a single circular two-dimensional Gaussian model to obtain the model parameters, the information of the model, such as flux, size, and degree of the major axis.
The model fitting was repeated until reduced $\chi^2$ approached one, and model fitting stops when the reduced $\chi^2$ does not decrease anymore.

Table\,\ref{table:image_para} lists the fitted parameters, including the observing frequencies, restoring beam size $B_{\rm maj,min}$, position angle $B_{\rm PA}$, total CLEAN flux density $S_{\rm KVN}$, peak flux density $S_{\rm p}$ in units of Jy per beam, rms in the image $\sigma$, the dynamic range (ratio of peak to rms) of the image D, and image quality factor $\xi_{\rm r}$.

\section{Results}\label{sec:result}
\subsection{Multifrequency light curves}\label{subsec:lc}

We present the results of the multifrequency radio observations of Mrk\,421.
Figure\,\ref{fig:lc} displays the multifrequency light curves of Mrk\,421 at 22, 43, and 86\,GHz observed between January 16, 2013 (MJD 56308) and June 1, 2018 (MJD 58270).
During the 5.5-year observation period, the flux density gradually decreases and several small flux enhancements are shown (which we call flares).
The light curves show a similar trend at all frequencies.
Although we can't show the 15\,GHz light curve of the OVRO in this study, the 15\,GHz light curve exhibits flux enhancements peaking at approximately MJD 56400, 56900, and 57200 \citep[see Figure\,1 in][]{Arbet-Engels+21}.
In the KVN light curves, only the flare at MJD 56900 is seen due to the relatively sparse data points. Because of that, the flare at MJD 56900 looks like a large flare. But the flare is very sharp indeed.

\startlongtable
\begin{deluxetable*}{llcccrrrrrr}
\tabletypesize{\small}
\tablecaption{Image parameters\label{table:image_para}}
\tablewidth{0pt}
\tablehead{
\colhead{Epoch} & 
\colhead{MJD} & 
\colhead{Band} & 
\colhead{$B_{\rm maj}$} &
\colhead{$B_{\rm min}$} & 
\colhead{$B_{\rm PA}$} & 
\colhead{$S_{\rm KVN}$} &
\colhead{$S_{\rm p}$} &
\colhead{$\sigma$} &
\colhead{$D$} &
\colhead{$\xi_{\rm r}$}\\ 
\colhead{} & 
\colhead{} & 
\colhead{} &
\colhead{(mas)} & 
\colhead{(mas)} & 
\colhead{($^\circ$)} &
\colhead{(Jy)} &
\colhead{(Jy/beam)} &
\colhead{(mJy/beam)} &
\colhead{} &
\colhead{}\\
\colhead{(1)} & 
\colhead{(2)} & 
\colhead{(3)} &
\colhead{(4)} & 
\colhead{(5)} & 
\colhead{(6)} &
\colhead{(7)} &
\colhead{(8)} &
\colhead{(9)} &
\colhead{(10)} &
\colhead{(11)}
}
\startdata
2013 Jan 16 &56308&K &0.300&0.189&$-$89.9&0.530&0.523&3&193&0.73\\
2013 Jan 16 &56308&Q &0.148&0.094&$-$85.2&0.508&0.496&6&89&0.72\\
2013 Jan 16 &56308&W &0.077&0.045&$-$72.6&0.278&0.281&11&26&0.75\\
2013 Apr 11 &56393&K &0.299&0.182&$-$82.7&0.435&0.435&4&103&0.79\\
2013 Apr 11 &56393&Q &0.147&0.096&$-$78.7&0.404&0.401&7&55&0.72\\
2013 Apr 11 &56393&W &0.076&0.043&$-$83.1&0.230&0.218&16&13&0.70\\
2013 May 07 &56419&K &0.300&0.180&$-$77.9&0.471&0.470&5&87&0.75\\
2013 May 07 &56419&Q &0.149&0.093&$-$77.2&0.408&0.404&11&35&0.68\\
2013 Oct 15 &56580&K &0.303&0.176&$-$77.2&0.507&0.505&5&95&0.82\\
2013 Oct 15 &56580&Q &0.151&0.089&$-$77.5&0.465&0.436&10&42&0.76\\
2013 Oct 15 &56580&W &0.077&0.043&$-$85.8&0.306&0.217&26&8&0.78\\
2013 Nov 19 &56615&K &0.304&0.179&$-$81.0&0.447&0.446&3&144&0.72\\
2013 Nov 19 &56615&Q &0.153&0.086&$-$84.2&0.390&0.369&5&79&0.80\\
2013 Nov 19 &56615&W &0.078&0.043&87.3&0.245&0.208&8&26&0.73\\
2013 Dec 24 &56650&K &0.283&0.170&79.6&0.476&0.472&6&84&0.74\\
2013 Dec 24 &56650&Q &0.146&0.086&79.5&0.471&0.446&7&65&0.83\\
2013 Dec 24 &56650&W &0.070&0.042&82.9&0.301&0.285&9&33&0.82\\
2014 Jan 27 &56684&K &0.298&0.187&$-$83.5&0.399&0.398&4&113&0.74\\
2014 Jan 27 &56684&Q &0.148&0.098&$-$79.6&0.378&0.368&5&75&0.71\\
2014 Feb 28 &56716&K &0.376&0.164&$-$63.9&0.394&0.393&5&76&0.72\\
2014 Feb 28 &56716&Q &0.176&0.086&$-$63.6&0.406&0.400&9&47&0.81\\
2014 Feb 28 &56716&W &0.113&0.037&$-$62.1&0.198&0.216&11&20&0.74\\
2014 Mar 22 &56738&K &0.309&0.172&$-$85.6&0.379&0.379&4&108&0.73\\
2014 Mar 22 &56738&Q &0.150&0.091&$-$85.7&0.353&0.353&6&61&0.83\\
2014 Mar 22 &56738&W &0.082&0.041&89.8&0.268&0.253&10&26&0.80\\
2014 Apr 22 &56769&K &0.304&0.175&$-$79.9&0.407&0.407&2&171&0.83\\
2014 Apr 22 &56769&Q &0.148&0.091&$-$78.4&0.389&0.384&3&131&0.79\\
2014 Apr 22 &56769&W &0.078&0.042&$-$79.7&0.302&0.286&8&37&0.85\\
2014 Sep 01 &56901&K &0.300&0.180&$-$80.1&0.723&0.720&7&99&0.83\\
2014 Sep 01 &56901&Q &0.142&0.095&$-$82.1&0.615&0.616&6&110&0.70\\
2014 Sep 01 &56901&W &0.078&0.043&$-$76.9&0.510&0.481&27&18&0.92\\
2014 Oct 29 &56959&K &0.330&0.160&86.2&0.430&0.428&5&79&0.74\\
2014 Oct 29 &56959&Q &0.174&0.075&$-$84.3&0.415&0.404&7&56&0.79\\
2014 Nov 28 &56989&K &0.334&0.164&$-$76.9&0.496&0.492&8&64&0.83\\
2014 Nov 28 &56989&Q &0.169&0.084&$-$70.2&0.509&0.508&9&57&0.70\\
2014 Dec 25 &57016&K &0.324&0.162&$-$84.6&0.464&0.462&6&78&0.87\\
2014 Dec 25 &57016&Q &0.148&0.090&$-$82.8&0.376&0.384&10&37&0.89\\
2015 Feb 23 &57076&K &0.312&0.170&$-$80.3&0.427&0.425&4&109&0.93\\
2015 Feb 23 &57076&Q &0.149&0.094&$-$78.8&0.368&0.365&3&115&0.77\\
2015 Feb 23 &57076&W &0.078&0.042&$-$77.7&0.233&0.220&7&34&0.84\\
2015 Mar 26 &57107&K &0.312&0.168&$-$84.2&0.398&0.396&4&111&0.98\\
2015 Mar 26 &57107&Q &0.147&0.091&$-$81.9&0.361&0.353&2&144&0.80\\
2015 Mar 26 &57107&W &0.079&0.041&$-$83.6&0.210&0.191&6&34&0.78\\
2015 Apr 30 &57142&K &0.323&0.163&$-$77.8&0.458&0.456&7&69&0.85\\
2015 Apr 30 &57142&Q &0.149&0.088&$-$81.9&0.407&0.404&4&110&0.84\\
2015 Apr 30 &57142&W &0.082&0.040&$-$78.0&0.304&0.288&11&25&0.79\\
2015 Sep 24 &57289&K &0.316&0.171&$-$69.6&0.528&0.528&6&95&0.84\\
2015 Sep 24 &57289&Q &0.156&0.083&$-$77.2&0.418&0.420&10&43&0.74\\
2015 Oct 23 &57318&K &0.347&0.165&80.8&0.438&0.436&9&46&0.81\\
2015 Nov 30 &57356&K &0.293&0.181&$-$86.9&0.396&0.396&4&96&0.89\\
2015 Nov 30 &57356&Q &0.141&0.096&$-$87.0&0.361&0.363&4&83&0.78\\
2015 Nov 30 &57356&W &0.076&0.043&$-$86.2&0.235&0.218&8&27&0.85\\
2015 Dec 28 &57384&K &0.300&0.175&$-$85.1&0.375&0.375&3&125&0.83\\
2015 Dec 28 &57384&Q &0.147&0.091&$-$81.1&0.344&0.343&2&144&0.78\\
2015 Dec 28 &57384&W &0.080&0.041&$-$85.3&0.217&0.202&6&36&0.79\\
2016 Jan 13 &57400&K &0.303&0.174&$-$88.6&0.337&0.337&2&139&0.71\\
2016 Jan 13 &57400&Q &0.147&0.091&$-$84.3&0.332&0.321&2&135&0.79\\
2016 Jan 13 &57400&W &0.080&0.041&89.5&0.191&0.164&6&26&0.84\\
2016 Mar 01 &57448&K &0.303&0.173&$-$85.4&0.417&0.417&3&123&0.90\\
2016 Mar 01 &57448&Q &0.149&0.089&$-$85.1&0.398&0.392&3&119&0.81\\
2016 Mar 01 &57448&W &0.077&0.042&$-$86.5&0.279&0.259&9&29&0.81\\
2016 Apr 24 &57502&K &0.301&0.185&$-$81.8&0.464&0.465&5&88&0.94\\
2016 Apr 24 &57502&Q &0.152&0.091&$-$80.9&0.427&0.410&3&117&0.90\\
2016 Apr 24 &57502&W &0.078&0.044&$-$83.8&0.245&0.251&6&44&0.77\\
2016 Oct 18 &57679&K &0.330&0.168&80.8&0.428&0.428&5&87&0.71\\
2016 Nov 27 &57719&K &0.310&0.183&$-$88.6&0.365&0.365&4&86&0.89\\
2016 Dec 28 &57750&K &0.290&0.184&$-$87.4&0.321&0.321&3&95&0.84\\
2016 Dec 28 &57750&Q &0.148&0.088&89.6&0.309&0.304&4&76&0.85\\
2017 Mar 28 &57840&K &0.372&0.172&$-$62.2&0.339&0.340&2&181&0.65\\
2017 Mar 28 &57840 &W &0.097 &0.042 &-61.7 &0.165 &0.149 &4 &37 &0.85\\
2017 Apr 19 &57862&K &0.310&0.170&$-$86.0&0.345&0.344&5&65&0.82\\
2017 Apr 19 &57862&Q &0.166&0.082&84.9&0.312&0.290&8&35&0.80\\
2017 May 21 &57894&K &0.316&0.174&83.9&0.341&0.341&5&68&0.78\\
2017 May 21 &57894&Q &0.168&0.081&80.3&0.295&0.282&4&68&0.75\\
2017 Jun 17 &57921&K &0.310&0.174&$-$77.3&0.356&0.355&5&75&0.93\\
2017 Jun 17 &57921&Q &0.164&0.080&$-$80.2&0.315&0.302&3&101&0.78\\
2017 Sep 19 &58015&K &0.321&0.180&$-$78.4&0.364&0.365&7&53&0.83\\
2017 Sep 19 &58015&Q &0.165&0.083&$-$78.9&0.312&0.308&3&98&0.80\\
2017 Oct 21 &58047&K &0.307&0.176&85.4&0.338&0.338&3&133&0.86\\
2017 Oct 21 &58047&Q &0.162&0.084&79.7&0.321&0.310&3&120&0.80\\
2017 Nov 04 &58061&K &0.330&0.185&$-$68.5&0.320&0.320&4&87&0.87\\
2017 Nov 04 &58061&Q &0.178&0.086&$-$69.0&0.310&0.301&4&79&0.77\\
2017 Nov 22 &58079&K &0.293&0.183&$-$84.4&0.299&0.298&4&82&0.85\\
2017 Nov 22 &58079&Q &0.156&0.085&$-$89.1&0.290&0.281&4&80&1.00\\
2017 Dec 06 &58093&K &0.296&0.178&$-$88.3&0.278&0.275&5&51&0.97\\
2018 Jan 13 &58131&Q &0.158&0.084&88.2&0.308&0.302&2&159&0.77\\
2018 Jan 13 &58131&W &0.075&0.043&$-$85.3&0.203&0.189&6&31&0.78\\
2018 Mar 25 &58202&K &0.300&0.184&$-$79.8&0.295&0.290&11&27&0.82\\
2018 Mar 25 &58202&W &0.085&0.042&$-$73.3&0.177&0.224&32&7&0.71\\
2018 Apr 19 &58227&K &0.307&0.174&$-$88.7&0.363&0.362&6&60&0.89\\
2018 Apr 19 &58227&Q &0.157&0.083&89.6&0.331&0.321&3&95&0.84\\
2018 Apr 19 &58227&W &0.082&0.040&88.8&0.239&0.244&11&22&0.95\\
2018 May 26 &58264&K &0.303&0.181&83.9&0.333&0.333&7&47&0.84\\
2018 May 31 &58269&K &0.331&0.171&80.0&0.327&0.327&8&41&0.79\\
\enddata
\tablecomments{ 
Column designation: 1~-~Date; 2~-~modified Julian date; 3~-~observing frequency band: 
K~-~22~GHz band; Q~-~43~GHz band; W~-~86~GHz band; D~-~129~GHz band; 
4-6~-~restoring beam:
4~-~major axis;
5~-~minor axis;
6~-~position angle of the major axis;
7~-~total CLEAN KVN flux density;
8~-~peak flux density;
9~-~off-source RMS in the image;
10~-~dynamic range of the image;
11~-~quality of the residual noise in the image (i.e., the ratio of the image root-mean-square noise to its mathematical expectation).
}
\end{deluxetable*}


\section{Analysis}\label{sec:analysis}

\subsection{Radio spectrum}\label{subsec:radio_sepc}
\begin{figure*}
\centering
\includegraphics[width=0.5\textwidth]{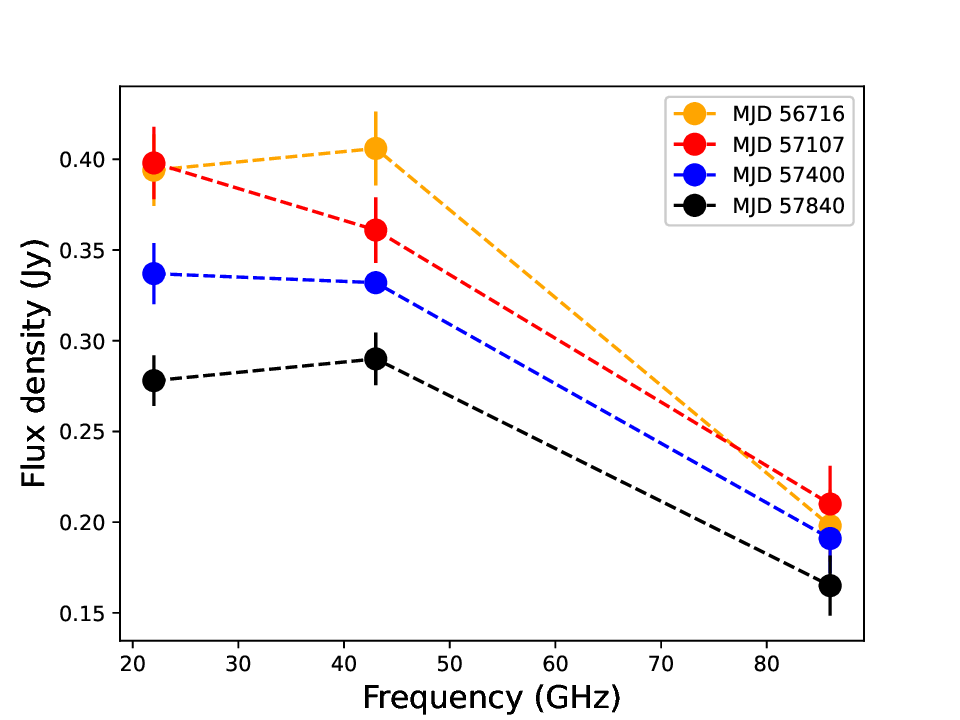}
\caption{\label{fig:local_minima} Plot of local minima on MJD 56716, 57107, 57400, and 57840 - 58093. }
\end{figure*}

\begin{figure}
\centering
\includegraphics[width=0.5\textwidth]{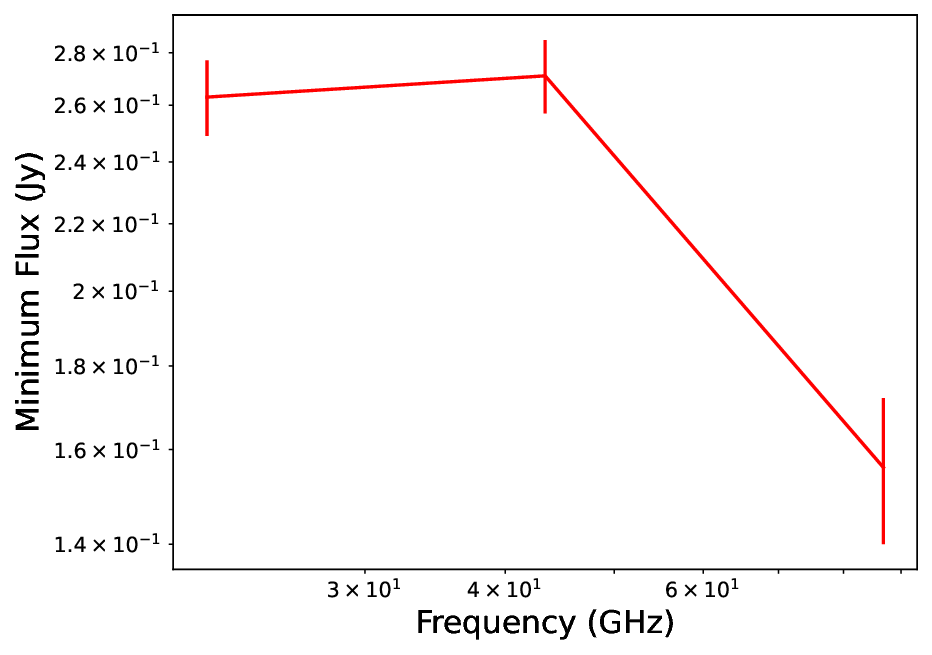}
\caption{\label{fig:min_fit_flux}The minimum flux density at each frequency, by fitting the four local minima obtained on  MJD 56716, 57107, 57400, and 57840 - 58093, and extrapolating each of the three frequencies to a common time (MJD 58269).} 
\end{figure}

The simultaneous multifrequency monitoring KVN observations enable us to investigate source spectra accurately without the time uncertainty. 
The multifrequency spectral analyses of variable blazars using arcsecond-scale observations by \cite{Kim+22} and \cite{Jeong+23} show that variable spectral properties can be revealed by subtracting off the quiescent spectrum in the multifrequency light curves. 
We calculated the quiescent spectrum by fitting the four local minima obtained on  MJD 56716, 57107, 57400, and 57840 - 58093 shown in Figure \ref{fig:local_minima}. The fourth local minimum we used is obtained on MJD 58093 at 22\,GHz, MJD 58079 at 43\,GHz, and MJD 57840 at 86\,GHz. The quiescent spectrum is obtained as 0.263$\pm${0.045}\,Jy, 0.271$\pm${0.021}\,Jy, and 0.156$\pm${0.027}\,Jy at 22, 43, and 86\,GHz, respectively, by extrapolating each of the three frequencies to a common time (MJD 58269), as shown in Figure \ref{fig:min_fit_flux}.

\begin{figure*}
\centering
\includegraphics[width=0.9\textwidth]{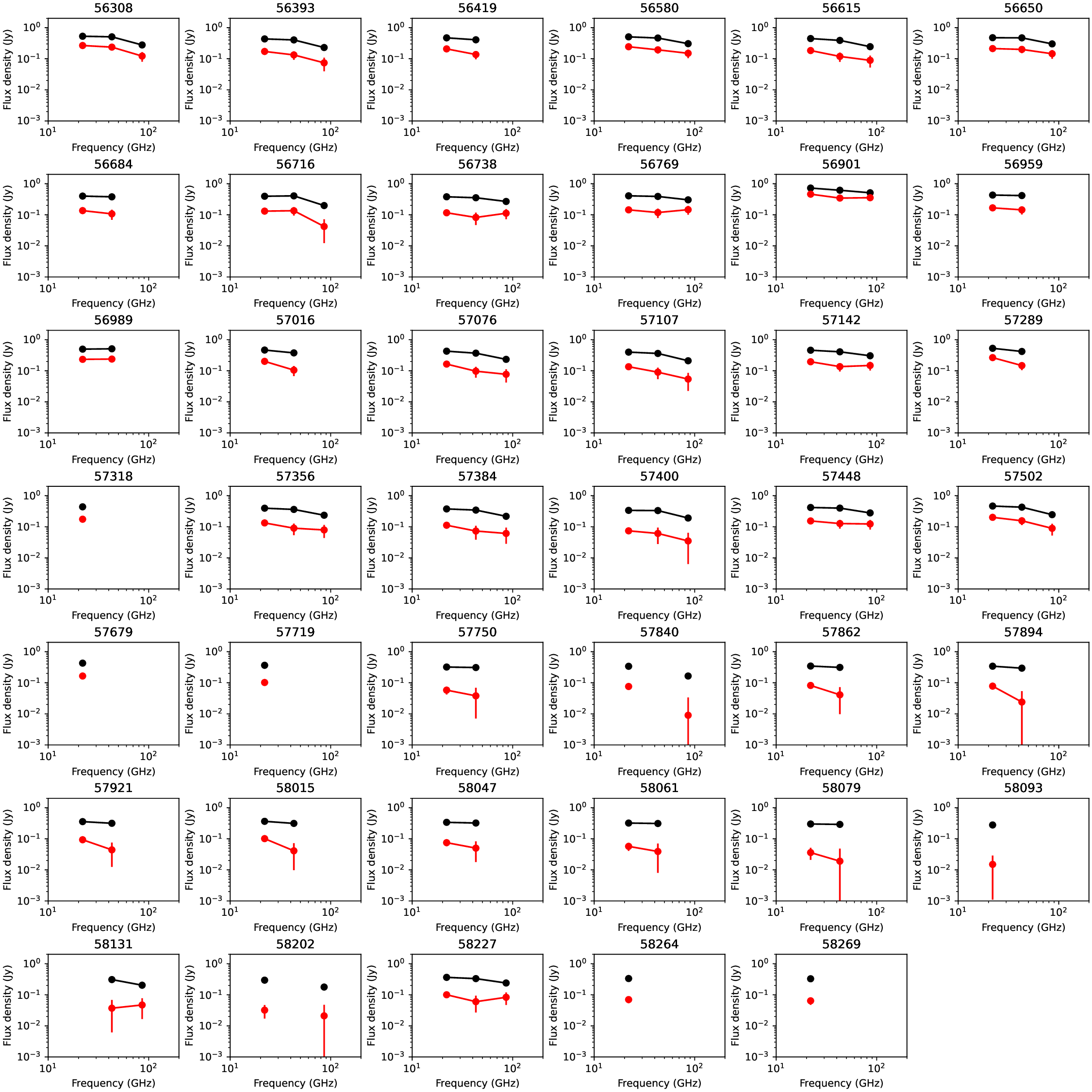}
\caption{\label{fig:spectrum}Spectrum of Mrk\,421. Black and red symbols represent the observed spectrum and quiescent-corrected spectrum, respectively.} 
\end{figure*}

The mas-scale spectra of Mrk\,421 are shown in Figure\,\ref{fig:spectrum}.
Out of the 41 epochs, data existed for at least two frequencies for 33 epochs. 
The observed spectrum is shown in black, and the quiescent-corrected spectrum (the observed spectrum by subtracting the quiescent spectrum) is shown in red.
The spectra generally appear to be flat between 22 and 43 GHz, and seem to be relatively steep between 43 and 86 GHz.
The quiescent-corrected spectra are sometimes consistent with the observed spectra and sometimes quite different.
Unlike the observed spectra, the quiescent-corrected spectra generally show inverted spectra between 22 and 43\,GHz. 
Although we were not able to precisely calculate the turnover frequency, $\nu_{\rm c}$, of Mrk\,421 in this analysis, we clearly find that the quiescent-corrected spectrum becomes steeper between 22 and 43\,GHz and flatter between 43 and 86\,GHz than the observed spectrum.
This implies that the quiescent emission correction for the source spectrum in the mas-scale may be important to study the intrinsic spectral properties of the variable emission regions.

\subsection{Spectral index}\label{subsec:sindex}

\begin{figure*}
\centering
\includegraphics[width=1\textwidth]{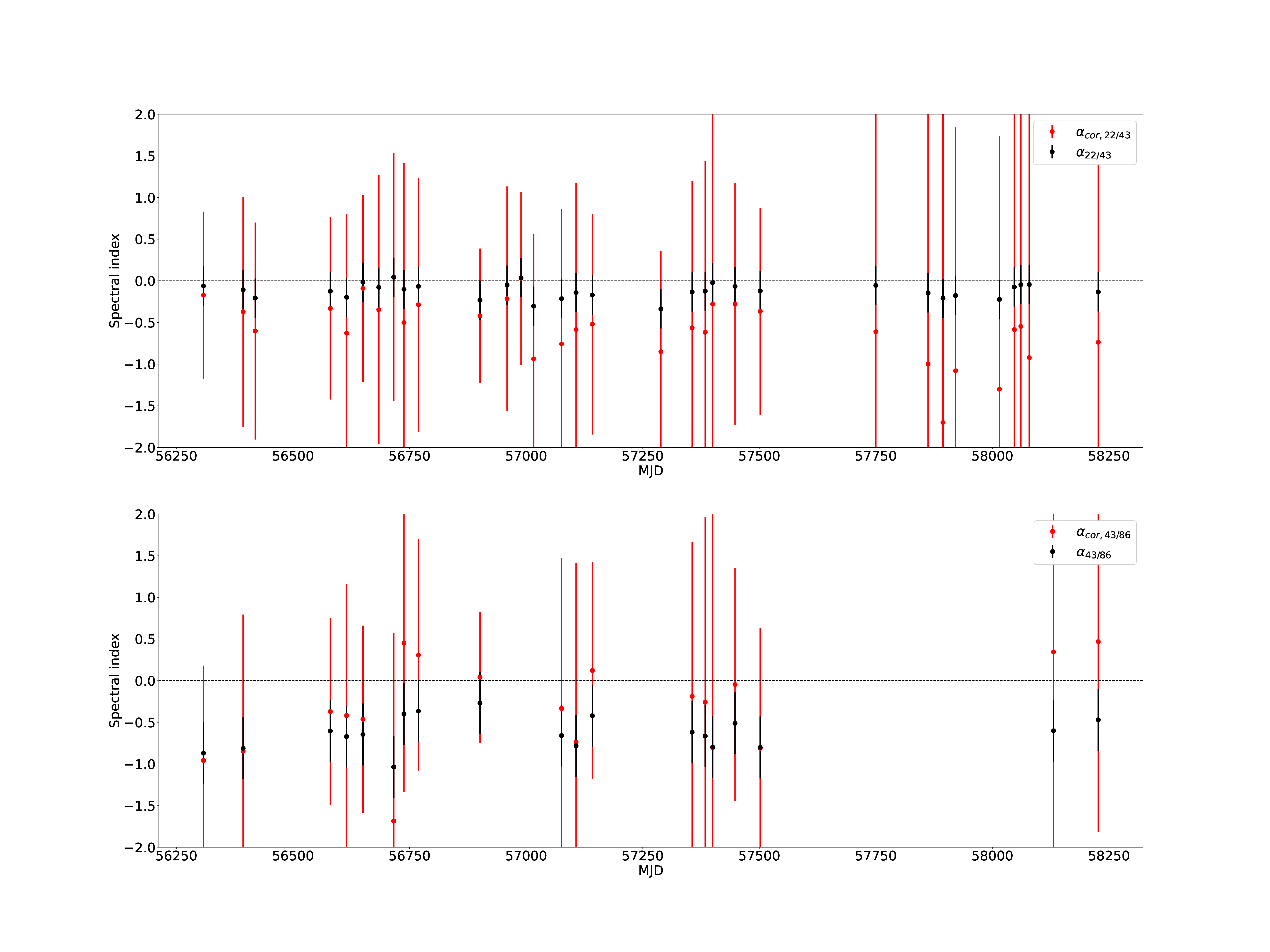}
\caption{\label{fig:sindex}Spectral index $\alpha$ of Mrk\,421. The top panel is $\alpha_{22/43}$ and the bottom panel is $\alpha_{43/86}$. The red symbols indicate the quiescent-corrected spectral index. A horizontal line is a guideline of zero.} 
\end{figure*}

The spectral index  $\alpha$ is defined as $S_{\nu}\propto\nu^{\alpha}$, where $\nu$ is the observing frequency, and $S_{\nu}$ is the flux density.
We calculated $\alpha$ for the data pairs 22 and 43\,GHz, $\alpha_{22/43}$, and 43 and 86\,GHz, $\alpha_{43/86}$~(see Figure~\ref{fig:sindex}).
The range of observed $\alpha_{22/43}$ was $-$0.34 - 0.04 with a mean $\alpha_{22/43}$ of $-$0.12, and $\alpha_{43/86}$ was $-$1.04 - $-$0.27 with a mean $\alpha_{43/86}$ of $-$0.63. 
The $\alpha_{22/43}$ was relatively flat, and $\alpha_{43/86}$ was steep.
As before, we calculated the quiescent-corrected spectral indices.
The range of the quiescent-corrected spectral indices between 22 and 43\, GHz $\alpha_{cor,22/43}$ was $-$1.7 - 0.04 with a mean $\alpha_{cor,22/43}$ of $-$0.57.
For 43 and 86\,GHz, the range of the quiescent-corrected spectral indices $\alpha_{cor,43/86}$ was $-$1.69 - $-$0.47 with a mean $\alpha_{cor,43/86}$ of $-$0.33. 

In Figure\,\ref{fig:sindex}, we see that the 22 and 43\,GHz quiescent-corrected spectral indices are consistently lower (steeper) than the observed spectral indices, although the errors are large. 
We noticed that the values of $\alpha_{22/43}$ were steeper  when the flares were around the peaks on MJD 56901 and MJD 57289, whereas those were flatter at local minima (MJD 56716 and MJD 57400).

\begin{figure*}
\centering
\includegraphics[width=1\textwidth]{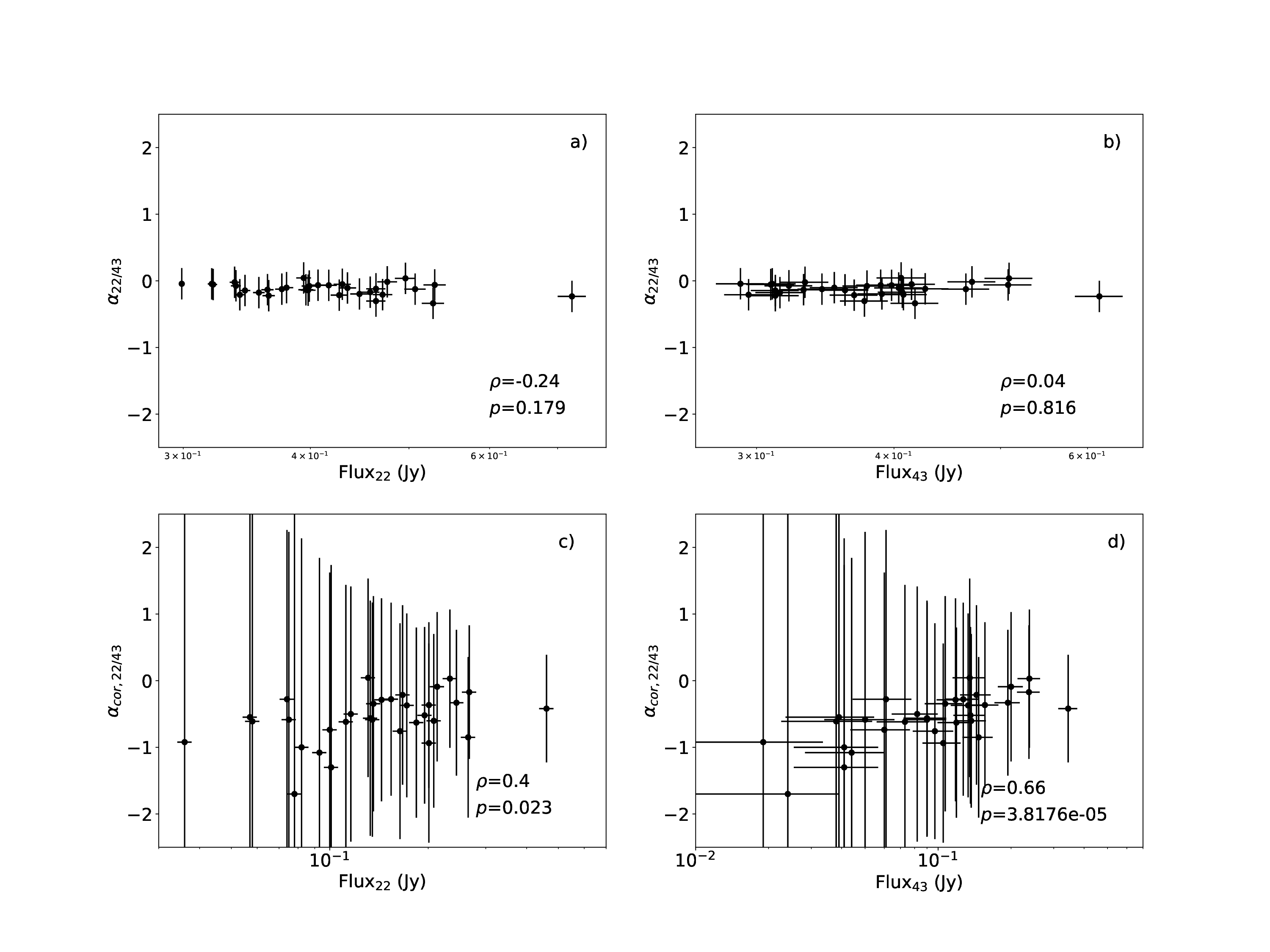}
\caption{\label{fig:sindex_flux_22_43}Spectral index as a function of the flux density. The left and right panels indicate 22 and 43\,GHz, respectively. a) and b) indicate $\alpha$ vs. flux density. c) and d) indicate the $\alpha_{cor}$ vs. flux density. $\rho$ and p values indicate coefficient and probability, respectively. } 
\end{figure*}

\begin{figure*}
\centering
\includegraphics[width=1\textwidth]{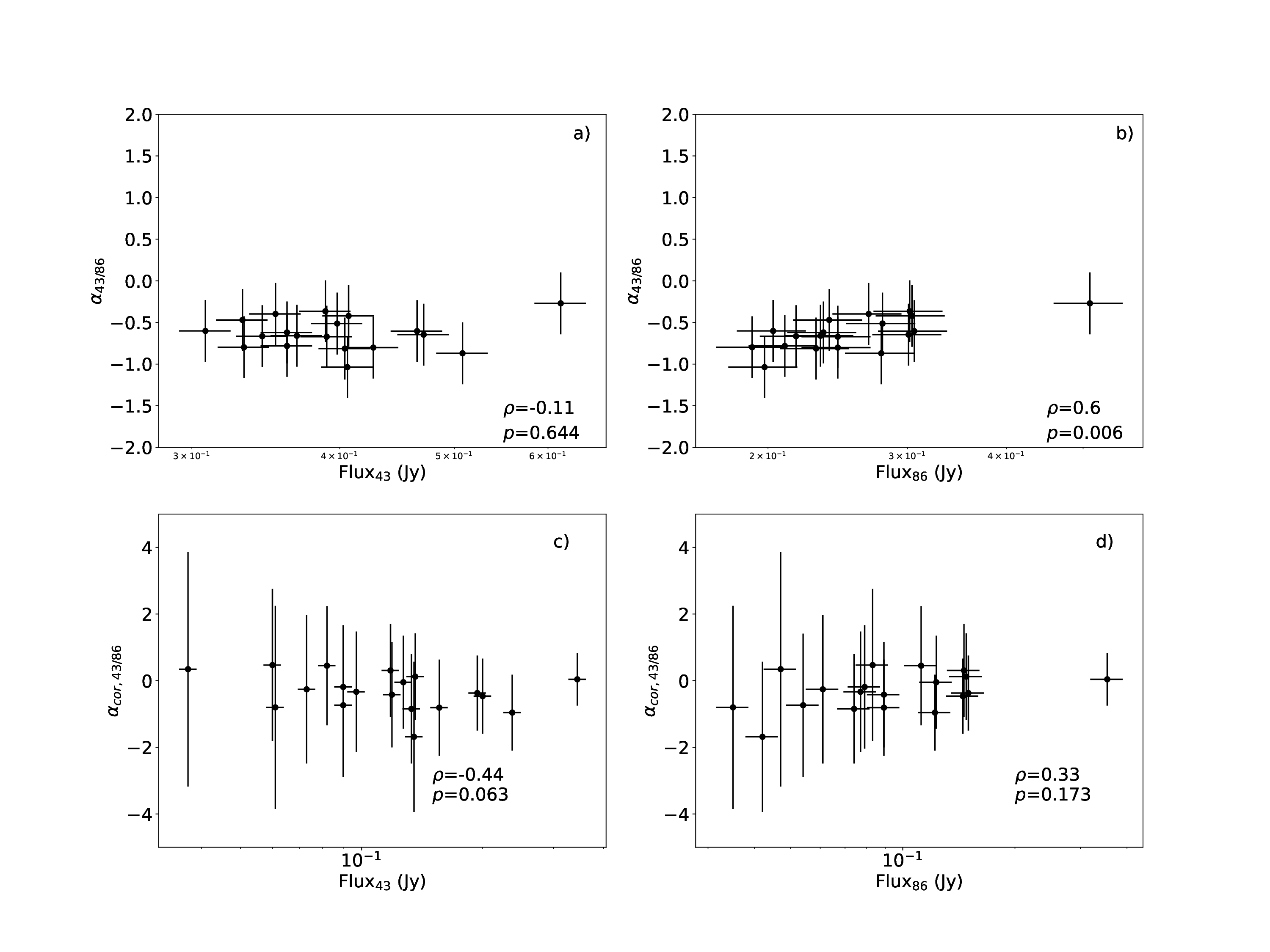}
\caption{\label{fig:sindex_flux_43_86}Spectral index as a function of the flux density. The left and right panels indicate 43 and 86\,GHz, respectively. a) and b) indicate $\alpha$ vs. flux density. c) and d) indicate the $\alpha_{cor}$ vs. flux density.  $\rho$ and p values indicate coefficient and probability, respectively. }
\end{figure*}

The 43 and 86\,GHz quiescent-corrected spectral indices showed more variability with spectral indices that were sometimes consistent, sometimes lower, and sometimes higher than the observed spectral indices, although the errors were large.
To further investigate the effect of correcting for the quiescent spectrum, we plotted the spectral indices versus flux density. In Figures\,\ref{fig:sindex_flux_22_43} - \ref{fig:sindex_flux_43_86}, a) and b) indicate the observed spectral index, and c) and d) indicate the quiescent-corrected spectral index versus flux density, respectively.
To analyze the relation between the spectral index and flux density quantitatively, we computed the Spearman correlation coefficient $\rho$ and probability $p$ for each case. 
The values of $\rho$ of $\alpha_{22/43}$ for 22\,GHz and 43\,GHz flux density were $-$0.24 and 0.04, respectively. 
The values of $\rho$ of $\alpha_{cor,22/43}$ for 22\,GHz and for 43\,GHz flux density were 0.4 and 0.66, respectively.. The values of $\rho$ of $\alpha_{43/86}$ for 43\,GHz and for 86\, GHz were $-$0.11 and 0.6, respectively.  The values of $\rho$ of $\alpha_{cor,43/86}$ for 43\,GHz and for 86\,GHz were $-$0.44 and 0.33, respectively. 

The original 22 and 43\,GHz spectral indices were uncorrelated with the flux densities.
For the quiescent-corrected spectral indices (c and d), the spectral indices and flux densities showed a significant correlation  (See, Figure\,\ref{fig:sindex_flux_22_43}).
In Figure\,\ref{fig:sindex_flux_43_86}, there is no correlation between $\alpha_{43/86}$ and the 43\,GHz flux density, but the $\alpha_{43/86}$ is correlated with 86\,GHz flux density.
The $\alpha_{43/86}$ increases with the 86\,GHz flux density.
 In the quiescent-corrected case, we now see an anti-correlation between the corrected spectral indices vs. 43\,GHz flux densities.
 The quiescent-corrected spectral indices were less correlated with the 86\,GHz flux densities than before.\\

\begin{figure*}
\centering
\includegraphics[width=1\textwidth]{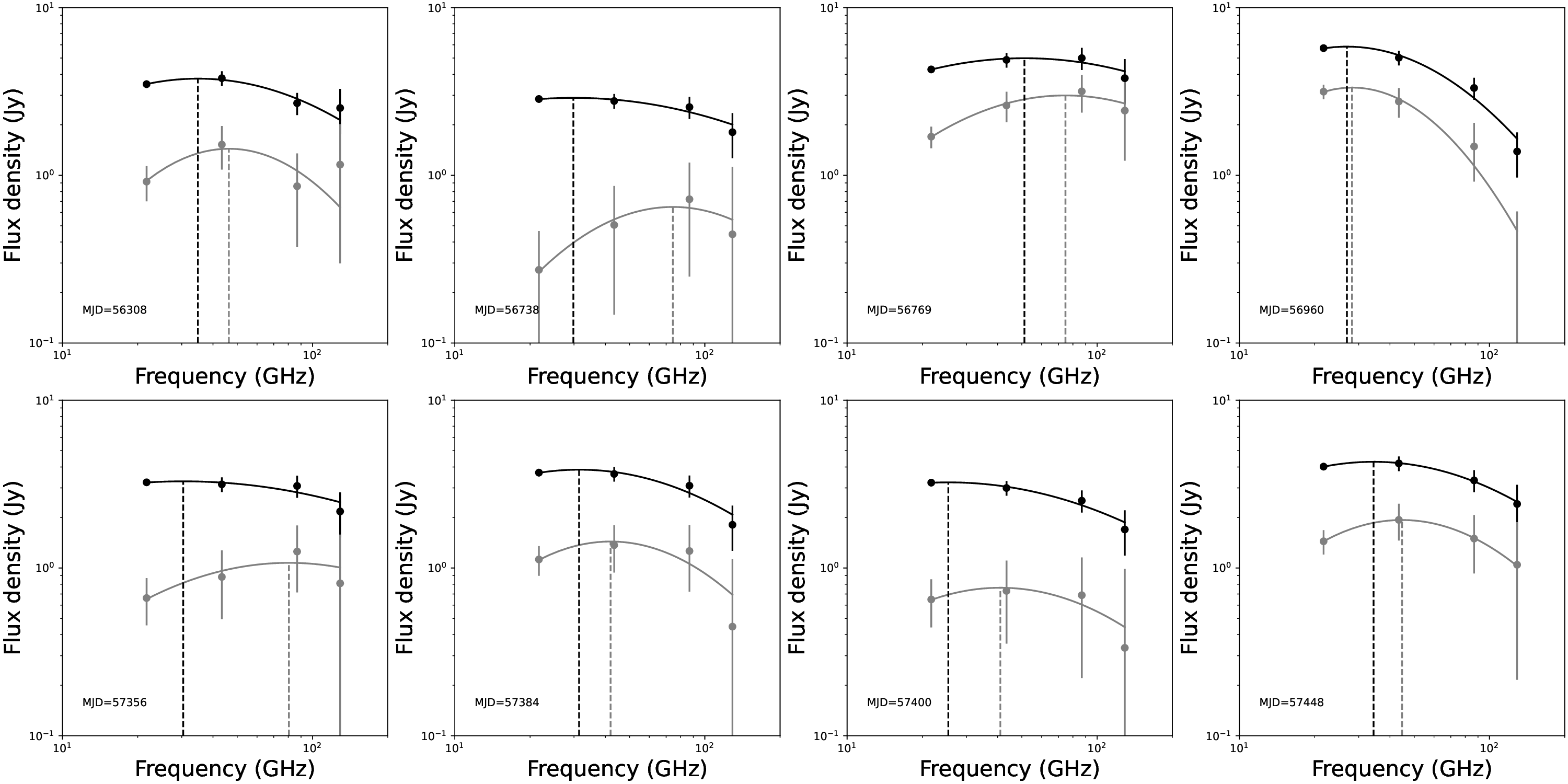}
\caption{\label{fig:oj_spectra}The spectra of OJ\,287. Black and grey symbols indicate the original and the quiescent-corrected spectra, respectively. Black and grey vertical lines correspond to the turnover frequency at the original spectrum and the quiescent-corrected spectrum, respectively.} 
\end{figure*}

\section{Discussion}\label{sec:discussion}

Figures \ref{fig:sindex_flux_22_43} and \ref{fig:sindex_flux_43_86} show that correcting for the quiescent spectrum can significantly affect the measurements of the spectral indices. 
This is of particular importance for calculations that require measurements of the synchrotron self-absorption frequency, for example, the magnetic field strength, $B_{\rm SSA} \propto \nu_{c}^{5}$.
This means that even small errors in measuring $\nu_{c}$ can lead to significant errors when measuring the magnetic field strength.
This suggests that magnetic field strength estimations based on multifrequency data should be carefully obtained with a proper determination of the optically thin quiescent spectrum of compact radio sources even on mas-scales.

To test the previous work, which did not consider correcting the quiescent spectrum, we re-analyzed the multifrequency spectral properties of OJ\,287 \citep{Lee+20}. The minimum flux densities of OJ\,287 were 2.58, 2.27, 1.83, and 1.36 at 22, 43, 86, and 129\,GHz, respectively. The spectral index between 86 and 129\,GHz was $-$0.73. We used this optically thin spectral index to correct the observed spectrum. 
We subtracted the minimum flux density obtained from the optically thin spectral index at each frequency for 8 epochs analyzed in \citet{Lee+20}. 
Figure\,\ref{fig:oj_spectra} shows the spectra of OJ\,287. The black and grey symbols indicate the observed and the quiescent-corrected spectra, respectively. After subtracting the optically thin spectrum, the spectrum was fitted with a curved power-law function \citep{Lee+16,Lee+17b,Lee+20}. Then, we obtained the turnover frequency and the peak flux density. The turnover frequencies obtained from the quiescent-corrected spectra were in the range of 28-80\,GHz, and the peak flux densities were 0.7-3.3\,Jy. The turnover frequencies were larger by a factor of approximately 2.5 larger than the ones obtained from the observed spectra. Because of the $B_{SSA} \propto \nu_{c}^{5}$, the synchrotron self-absorption magnetic field strength might be underestimated in \citet{Lee+20}. 
This could also potentially change the comparison result between the equipartition magnetic field strength and synchrotron self-absorption magnetic field strength.

\section{Summary}\label{sec:summary}
We have presented the results of simultaneous multifrequency observations at 22, 43, and 86\,GHz with the KVN.
The light curves showed a global decaying and several small radio flares.
To separate the variable component and the quiescent component in the light curves, we subtracted the quiescent component, assuming the minimum flux density.
We found that the observed spectra were flat at 22-43\,GHz, and relatively steep at 43-86~GHz, whereas the quiescent-corrected spectra are sometimes quite different from the observed spectra (e.g., sometimes inverted at 22-43\,GHz ).
This means that the turnover frequency, $\nu_{c}$ is shifted to a higher frequency.  Based on the assumption of the synchrotron self-absorption magnetic field strength estimation,  $B_{\rm SSA} \propto \nu_{c}^5$.
Therefore, to properly estimate the magnetic field strength, subtracting the quiescent spectrum in the KVN light curves is important.

\acknowledgments
We thank the anonymous reviewer for valuable comments and suggestions that helped to improve the paper. We are grateful to all staff members in KVN who helped to operate the array and to correlate the data. The KVN is a facility operated by the Korea Astronomy and Space Science Institute. The KVN operations are supported by KREONET (Korea Research Environment Open NETwork) which is managed and operated by KISTI (Korea Institute of Science and Technology Information). This work was supported by the National Research Foundation of Korea (NRF) grant funded by the Korea government (MIST) (2020R1A2C2009003).


\begin{thebibliography}{}
\bibitem[Acciari et al.(2011)]{Acciari+11} Acciari, V.~A., Aliu, E., Arlen, T., et al.\ 2011, \apj, 738, 25. doi:10.1088/0004-637X/738/1/25
\bibitem[Aleksi{\'c} et al.(2015)]{Aleksic+15} Aleksi{\'c}, J., Ansoldi, S., Antonelli, L.~A., et al.\ 2015, \aap, 576, A126. doi:10.1051/0004-6361/201424216
\bibitem[Algaba et al.(2015)]{Algaba+15} Algaba, J.-C., Zhao, G.-Y., Lee, S.-S., et al.\ 2015, Journal of Korean Astronomical Society, 48, 237. doi:10.5303/JKAS.2015.48.5.237
\bibitem[Algaba et al.(2018)]{Algaba+18a} Algaba, J.-C., Lee, S.-S., Kim, D.-W., et al.\ 2018, \apj, 852, 30. doi:10.3847/1538-4357/aa9e50
\bibitem[Arbet-Engels et al.(2021)]{Arbet-Engels+21} Arbet-Engels, A., Baack, D., Balbo, M., et al.\ 2021, \aap, 647, A88. doi:10.1051/0004-6361/201935557
\bibitem[Charlot et al.(2006)]{Charlot+06} Charlot, P., Gabuzda, D.~C., Sol, H., et al.\ 2006, \aap, 457, 455. doi:10.1051/0004-6361:20054078
\bibitem[Deller et al.(2007)]{Deller+07} Deller, A.~T., Tingay, S.~J., Bailes, M., et al.\ 2007, \pasp, 119, 318. doi:10.1086/513572
\bibitem[Hodgson et al.(2016)]{Hodgson+16} Hodgson, J., Lee, S.-S., Zhao, G.-Y., et al. 2016, JKAS, 49, 137
\bibitem[Horan et al.(2009)]{Horan+09} Horan, D., Acciari, V.~A., Bradbury, S.~M., et al.\ 2009, \apj, 695, 596. doi:10.1088/0004-637X/695/1/596
\bibitem[Jeong et al.(2023)]{Jeong+23} Jeong, H.-W., Lee, S.-S., Cheong, W.~Y., et al.\ 2023, \mnras, 523, 5703. doi:10.1093/mnras/stad1736
\bibitem[Kang et al.(2021)]{Kang+21} Kang, S., Lee, S.-S., Hodgson, J., et al.\ 2021, \aap, 651, A74. doi:10.1051/0004-6361/202040198
\bibitem[Katarzy{\'n}ski et al.(2003)]{Katarzynski+03} Katarzy{\'n}ski, K., Sol, H., \& Kus, A.\ 2003, \aap, 410, 101. doi:10.1051/0004-6361:20031245
\bibitem[Kim et al.(2022)]{Kim+22} Kim, S.-H., Lee, S.-S., Lee, J.~W., et al.\ 2022, \mnras, 510, 815. doi:10.1093/mnras/stab3473
\bibitem[Lee et al.(2015)]{Lee+15} Lee, S.-S., Byun, D.-Y., Oh, C.~S., et al.\ 2015, Journal of Korean Astronomical Society, 48, 229. doi:10.5303/JKAS.2015.48.5.229
\bibitem[Lee et al.(2016)]{Lee+16} Lee, S.-S., Wajima, K., Algaba, J.-C., et al.\ 2016, \apjs, 227, 8 
\bibitem[Lee et al.(2017a)]{Lee+17a} Lee, J.~W., Lee, S.-S., Hodgson, J.~A., et al.\ 2017a, \apj, 841, 119 
\bibitem[Lee et al.(2017b)]{Lee+17b} Lee, J.~W., Sohn, B.~W., Byun, D.-Y., Lee, J.~A., \& Kim, S.~S.\ 2017b, \aap, 601, A12
\bibitem[Lee et al.(2020)]{Lee+20} Lee, J.~W., Lee, S.-S., Algaba, J.-C., et al.\ 2020, \apj, 902, 104. doi:10.3847/1538-4357/abb4e5
\bibitem[Lobanov et al.(2006)]{Lobanov+06} Lobanov, A.~P., Krichbaum, T.~P., Witzel, A., et al.\ 2006, \pasj, 58, 253. doi:10.1093/pasj/58.2.253
\bibitem[Punch et al.(1992)]{Punch+92} Punch, M., Akerlof, C.~W., Cawley, M.~F., et al.\ 1992, \nat, 358, 477. doi:10.1038/358477a0
\bibitem[Rioja \& Dodson(2011)]{Rioja+11} Rioja, M. \& Dodson, R.\ 2011, \aj, 141, 114. doi:10.1088/0004-6256/141/4/114
\bibitem[Ulrich et al.(1997)]{Ulrich+97} Ulrich, M.-H., Maraschi, L., \& Urry, C.~M.\ 1997, \araa, 35, 445. doi:10.1146/annurev.astro.35.1.445
\bibitem[Urry \& Padovani(1995)]{Urry+95} Urry, C.~M. \& Padovani, P.\ 1995, \pasp, 107, 803. doi:10.1086/133630
\bibitem[Zhu et al.(2016)]{Zhu+16} Zhu, Q., Yan, D., Zhang, P., et al.\ 2016, \mnras, 463, 4481. doi:10.1093/mnras/stw2346
\end{thebibliography}
\end{document}